\documentclass[10pt,conference]{IEEEtran}
\IEEEoverridecommandlockouts

\usepackage[cmex10]{amsmath}
\usepackage{amssymb,color,graphicx,epsfig,upgreek,psfrag}

\usepackage{cite}
\usepackage[tight,footnotesize]{subfigure}
\newcommand{\sE}{\mathbb{E}}		% expectation operator
\newcommand{\sCN}{\mathcal{CN}}		% complex Gaussian distribution
\newcommand{\seye}{\mathbf{I}}		% eye matrix
\newcommand{\sdiag}{\text{diag}}   % diagonal matrix
\newcommand{\define}{\triangleq}

\newcommand{\su}{\text{u}}

\newcommand{\si} {i}			% Relay index
\newcommand{\sk} {k}			% update index
\newcommand{\st} {n}			% time index

\newcommand{\sRset} {\mathcal{R}}			% set of Relays
\newcommand{\sSset} {\mathcal{S}}			% Source
\newcommand{\sDset} {\mathcal{D}}			% Destination

\newcommand{\sT} {T}			% block length
\newcommand{\sTone} {\mathcal{\sT}_\text{p}^{+}}			% training interval 1
\newcommand{\sTtwo} {\mathcal{\sT}_\text{p}^{-}}			% training interval 2
\newcommand{\sTpset} {\mathcal{\sT_\text{p}}}			% training interval 
\newcommand{\sTdset} {\mathcal{\sT_\text{d}}}			% training interval 

\newcommand{\sR} {R}			% number of relays
\newcommand{\sRmax} {\sR_\text{max}}			% number of relays

\newcommand{\sh} {h}			% backward channel
\newcommand{\shi} {\sh_\si}

\newcommand{\shbari} {\bar{\sh}_\si}	
\newcommand{\shbar} {\bar{\sh}}	
\newcommand{\svhbar} {\mathbf{\bar{\sh}}}
\newcommand{\sg} {g}			% forward channel
\newcommand{\sgi} {\sg_\si}
	
\newcommand{\sgbari} {\bar{\sg}_\si}	
\newcommand{\sgbar} {\bar{\sg}}	

\newcommand{\sGbar} {\mathbf{\bar{G}}}
\newcommand{\shcomp} {\xi}
\newcommand{\sucomp} {\eta}

\newcommand{\ssymb} {s}			% symbol
\newcommand{\spilot} {\ssymb_\text{p}}			% pilots
\newcommand{\sn} {w}			% noise at relays
\newcommand{\svn} {\mathbf{\sn}}			% noise at relays
\newcommand{\sni} {\sn_\si}		% noise at relays
\newcommand{\sv} {v}			% noise at destination
\newcommand{\sNo} {N_0}			% noise variance

\newcommand{\sx} {x}			% received signal at relays
\newcommand{\sxi} {\sx_\si}

\newcommand{\sr} {r}			% transmit signal at relays
\newcommand{\sri} {\sr_\si}	

\newcommand{\spnorm} {\lambda_\si}	

\newcommand{\sw} {\alpha}				% relay weights
\newcommand{\swi} {\sw_{\si}}
\newcommand{\svw} {\boldsymbol{\sw}}
\newcommand{\svwd} {\svw_{\sk}}
\newcommand{\swt} {\tilde\sw_{\sk,\si}}
\newcommand{\svwt} {\tilde\svw_{\sk}}
\newcommand{\swtu} {\tilde\sw_{\sk,\si}^\prime}
\newcommand{\svwtu} {\tilde\svw_{\sk}^\prime}

\newcommand{\svwta} {\svwt^{+}}
\newcommand{\svwtb} {\svwt^{-}}
\newcommand{\svwk} {\svw_\sk}
\newcommand{\svwkp} {\svw_{\sk+1}}
\newcommand{\sy} {y}			% received signal

\newcommand{\sq} {q}				% perturbation set

\newcommand{\svq} {\mathbf{\sq}}

\newcommand{\sQset} {\mathbf{Q}} 
\newcommand{\sQ} {\mathbf{F}}  
\newcommand{\sN} {N}				% no of perturbation vectors

\newcommand{\sfb} {c}			% feedback signal
\newcommand{\sfbk} {\sfb_\sk}			% feedback signal
\newcommand{\sbeta} {\mu}	% step coefficient
\newcommand{\sfbm} {\sJ}		% feedback measure
\newcommand{\sfbmkp} {\sfbm_{\sk+1}}		% feedback measure
\newcommand{\sfbmone} {\tilde\sfbm_\sk}		% feedback measure
\newcommand{\sfbmonep} {\sfbmone^{+}}		% feedback measure
\newcommand{\sfbmonem} {\sfbmone^{-}}		% feedback measure
\newcommand{\sfbmtwo} {\sfbm_{\sk}}		% feedback measure

\newcommand{\sJ} {\gamma}			% cost function

\newcommand{\ssnr} {\rho}		% SNR
\newcommand{\ssnrk} {\ssnr_\sk}		% SNR

\newcommand{\sP} {P}    % general power 
\newcommand{\sPs} {\sP_\ssymb}   % source transmit power
\newcommand{\sPsum}{\Delta}		% relay transmit sum power
\renewcommand{\sPsum}{\bar{P}}		% relay transmit sum power

\newcommand{\sPrx} {\sP_{\sDset}}		% received signal power
\newcommand{\sPrxhat} {\hat{\sP}_{\sDset}}		% received signal power
\newcommand{\ssnrhat} {\hat{\ssnr}}
\newcommand{\shcomphat} {\hat{\shcomp}}
\newcommand{\sTsets} {\mathcal{\sT'}}

\graphicspath{{plots/}}

\begin{document}
% \title{Distributed Beamforming for Relay Systems\\ based on Perturbation }

\title{Perturbation-based Distributed Beamforming for\\[-1mm] Wireless Relay Networks}

\author{%
~\\[-7mm]
\IEEEauthorblockN{Peter Fertl\IEEEauthorrefmark{1}, Ari Hottinen\IEEEauthorrefmark{2}, and Gerald Matz\IEEEauthorrefmark{1}\thanks{This work was supported by the STREP project MASCOT
(IST-026905) within the Sixth Framework Programme of the European Commission.}\\[1.5mm]
}\IEEEauthorblockA{\IEEEauthorrefmark{1}\small %$^{1)}$
Institute of Communications and Radio-Frequency Engineering, Vienna University of Technology\\[-.5mm]
 Gusshausstrasse 25/389, A-1040 Vienna, Austria, phone: +43 1 58801 38942, email: pfertl@nt.tuwien.ac.at\\[.5mm]}
\IEEEauthorblockA{\IEEEauthorrefmark{2}\small Nokia Research Center, P.O. Box 407, FIN-00045 Nokia Group, Finland, email: ari.hottinen@nokia.com\\[-1mm]}}%%
%
%\IEEEauthorblockN{Peter Fertl and Gerald Matz\thanks{This work was supported by the STREP project MASCOT
%  (IST-026905) within the Sixth Framework Programme of the European Commission and by the WWTF project MOHAWI (MA 44).}}
%\IEEEauthorblockA{ Institute of Communications and Radio-Frequency Engineering\\ Vienna University of Technology\\
% Gusshausstrasse 25/389, A-1040 Vienna, Austria \\
% Phone: +43 1 58801 38942, Email: pfertl@nt.tuwien.ac.at}
% \and
% \IEEEauthorblockN{Ari Hottinen}
%\IEEEauthorblockA{Radio Communications CTC\\
%Nokia Research Center\\
%P.O. Box 407, FIN-00045 Nokia Group, Finland\\
%Email: ari.hottinen@nokia.com}}


\maketitle

%% **************** ABSTRACT ***************************
\begin{abstract}
%\boldmath
This paper deals with distributed beamforming %and power allocation 
techniques for wireless networks with half-duplex amplify-and-forward relays. 
Existing schemes optimize the beamforming weights based on the assumption that
channel state information (CSI) is available at the relays.
% require the knowledge of all, or at least, the local channels at the relays to calculate the corresponding beamforming weights. 
We propose to use adaptive beamforming based on deterministic perturbations 
% to achieve the same performance gains with only 
and limited feedback (1-bit) from the destination to the relays in order
to avoid CSI at the relays. Two scalable perturbation schemes are considered and
practical implementation aspects are addressed.
% The proposed adaptive protocol allows to cope with the scalability of such networks. 
% Moreover, we compare various perturbation schemes and analyze their convergence properties. 
Simulation results confirm that the proposed techniques closely approach optimum performance
% the performance gains achievable with optimal power allocation methods and allow for 
and have satisfactory tracking properties in time-varying environments. % genauer techniques (deterministic - stochastic weights, power-snr measure), scalable? 
\end{abstract}

% **************** INTRODUCTION ***************************

\section{Introduction}

\vspace*{-.5mm}
\subsection{Background}

\vspace*{-.5mm}
% Spatial diversity offers significant performance enhancement in multiple antenna systems by mitigating fading and improving link reliability. 
Terminal cooperation in wireless networks has been recognized as a means
to form virtual arrays that can realize spatial diversity in a distributed fashion.
% In such networks the nodes cooperate to form virtual arrays, thus realizing spatial diversity in a distributed fashion. %Under this context several cooperation protocols have been analyzed \cite{sendonaris03,rohit_jsac03,laneman_2004a}. 
An important special case is distributed beamforming with half-duplex amplify-and-forward (AF) relays.
%  networks have become increasingly popular. %and, in particular, address the problem of optimal power control at the relays. 
% short description of BF & PA schemes:
The coherent AF scheme in \cite{dana03} requires local channel phase information at the relays to achieve coherent phase combining with equal power at all relays. Beamforming with non-uniform power allocation (PA) 
under a sum power constraint \cite{Larsson:2003aa,Hammerstrom:2004aa}
and under individual relay power constraints \cite{Jing:2007aa}
offers significant performance gains. 
% PA under the assumption of individual power constraints at the relays was studied in \cite{Jing:2007aa}. 
However, optimal beamforming with PA places strong requirements regarding channel state information (CSI) at the relays.
%requires that each relay knows {\em all} channels perfectly.
%  or, if only local channel state information (CSI) is available, feedback from the destination. 
For centralized arrays with co-located antennas, 
this requirement has been circumvented
% short desc. of perturbation in centralized MISO/multiuser systems: 
% For multiple-antenna systems with centralized arrays the problem of required channel knowledge at the transmitter can be solved 
by adaptive gradient beamforming techniques 
% (also known as {\em subspace trackin\gammag}) 
% (cf.\Ê\cite{Banister:2003aa,Banister:2005aa} for the centralized case). 
that iteratively adjust the beamforming weights using stochastic vector perturbations and limited feedback from the 
destination \cite{Banister:2003aa}. 
A related approach based on deterministic perturbations is presented in \cite{Raghothaman:2003aa}. 
%An extension of these schemes to the relay setup with PA under an sum power constraint is not straightforward since each relay has to know the weights of all other relays. 
% Quantized feedback \cite{Nguyen:2006aa}
In a similar spirit, feedback-assisted distributed beamforming with phase perturbation in wireless networks was considered in \cite{Mudumbai:2006aa} and extended to the multiuser context in \cite{Thukral:2007ab}. However, both methods do not assume a relay setup and do not address distributed PA.
%  the problem of power control which is a crucial issue in relay networks to prevent possible noise amplification.

\subsection{Contribution and Organization of Paper}
We consider perturbation-based beamforming (PB-BF) with 1-bit feedback in a relay network. 
Under the assumption of a sum power constraint, the relays use the feedback bit to adapt their beamforming weights 
% (i.e., phase and power) 
% to current fading, noise, and path loss situation, such that the 
in order to maximize either the signal-to-noise ratio (SNR) or the received signal power at the destination. 
% We note that t
This approach does not require any CSI at the relays. % Within this context, 
Two different perturbation schemes are investigated, both of which are based on
deterministic perturbation sets to avoid extensive signaling/feedback overhead. % between the nodes. 
Within this context, we present a scalable protocol, discuss implementation aspects, and provide numerical performance comparisons. 
% with the beamforming methods in \cite{dana03,Larsson:2003aa,Jing:2007aa}. 
Simulation results corroborate that our approach can satisfactorily track time-varying channels in non-static environments.
We note that in the context of wireless ad-hoc networks a related idea was touched upon in \cite{Li:2008aa} without explicitly addressing the important practical problem of weight exchange.
%, however, assuming perfect exchange of the beamforming weights among the relays.
% however, it was assumed that the nodes can perfectly share their weights. % perfectlythe important subject of weight sharing between the relays was totally ignored. 

The rest of the paper is organized as follows. Section \ref{sec:system} introduces the system model and Section \ref{sec:perturb} proposes perturbation-based distributed beamforming with 1-bit feedback. %Implementation aspects are addressed in Section \ref{sec:implementation} and 
A comparison with optimum batch solutions is provided in Section \ref{sec:PA}. Section \ref{sec:results} discusses simulation results and conclusions are provided in Section \ref{sec:conclusion}.

%{\em Notation:}     %$\sCN(\mu,\sigma^2)$ denotes the complex normal distribution with mean $\mu$ and variance $\sigma^2$.
%
%{\em Define:} SNR

% ************** SYTEM MODEL *********************************

\section{System Model}

\label{sec:system}
\begin{figure}
\centering
\vspace*{2mm}
\input{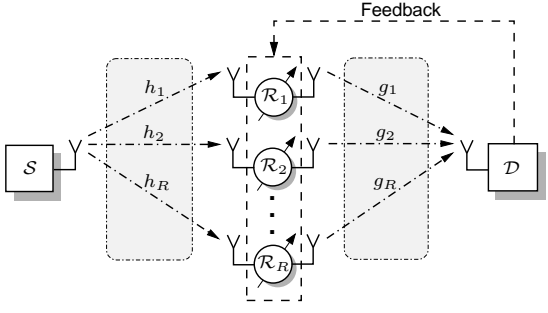}
%\includegraphics{plots/relay_system_fb4.eps}
%\vspace*{-1mm}
\caption{Wireless relay network with feedback.}
\vspace*{-1mm}
\label{fig:relay}
\vspace*{-2mm}
\end{figure}

We consider a perfectly synchronized wireless network with single antenna nodes where a single source $\sSset$ communicates with a single destination $\sDset$ via $\sR$ half-duplex relays $\sRset_\si$, $\si=1,\dots,\sR$ (cf.\ Fig.\,\ref{fig:relay}). 
The half-duplex constraint necessitates a two-hop protocol. % spanning two time slots.
%
% single antenna nodes where one source $\sSset$ communicates with one destination $\sDset$ through a set of $\sR$ half-duplex relay terminals (denoted by $\sRset$), as depicted in Fig.\,\ref{fig:relay}. Throughout this work, we assume that transmission/reception between the nodes is perfectly synchronized and that there is no direct link between $\sSset$ and $\sDset$. %We furthermore assume frequency-flat block fading for all channels in the network with block length $\sN$. 
%In the following, we denote a channel between the source and a relay as backward channel and a channel between a relay and the destination as forward channel.
% With regard to half-duplex (i.e., the nodes cannot transmit and receive simultaneously), transmission takes place in two hops over two time intervals. For simplicity, we focus here on symbol-wise transmission: 
%
In the first hop, $\sSset$ transmits the signal $\sqrt{\sPs}\ssymb$
to the relays which receive
\vspace*{-.5mm}
\begin{equation}
\sxi = \sqrt{\sPs}\shi\ssymb + \sni, \quad i=1,\dots,\sR \,. \label{eq:xi}
\vspace*{-.5mm}
\end{equation}
Here, $\ssymb$ is the transmit symbol normalized as $\sE\{|\ssymb|^2\}\!=\!1$ ($\sE\{\cdot\}$ denotes expectation), 
%is chosen equally likely from a symbol alphabet $\sA$, 
$\sPs$ denotes the average transmit power of $\sSset$,
$\shi$ is the complex coefficient of the flat fading ``backward'' channel\footnote{Note that our discussion does not presume specific channel statistics.} 
between $\sSset$ and $\sRset_\si$, 
and $\sni\!\sim\!\sCN(0,\sNo)$ denotes i.i.d.\ complex Gaussian noise.
In the AF scenario considered, the second hop amounts to each relay transmitting a complex scaled version of the signal it has received, i.e., 
\begin{equation} %\swi \frac{\sxi}{\sqrt{\sE\{|\sxi|^2\}}}=
\sri = \swi^* \spnorm\, \sxi \,,\quad\text{with }\,
\spnorm \define \sqrt{\!\frac{\sP}{\sPs |\shi|^2 + \sNo}}. \label{eq:asi}
\end{equation}
Here, complex conjugation (superscript $^*$) of the beamforming weights $\swi$ will simplify notation later on, and $\spnorm$ is a power normalization factor such that the average relay power is $\sE\{|\sri|^2|\shi\}\!=\!|\swi|^2\sP$.
%In the AF scenario considered, the second hop amounts , each relay normalizes the received signal such that the average per-relay transmit power $\sE\{|\sri|^2|\shi\}\!=\!\sP$ and applies the complex beamforming weight $\swi$, i.e. % the second hop amounts to each relay transmitting a complex scaled version of the signal it has received, i.e. 
%\begin{equation} %\swi \frac{\sxi}{\sqrt{\sE\{|\sxi|^2\}}}=
%\sri = \swi^* \spnorm\, \sxi \,,\quad\text{with }\,
%\spnorm \define \sqrt{\!\frac{\sP}{\sPs |\shi|^2 + \sNo}}; \label{eq:asi}
%\end{equation}
%here, complex conjugation (superscript $^*$) of $\swi$ simplifies notation later on. %, and $\spnorm$ is a power normalization factor .
The destination receives
$%\begin{equation}\label{eq:y1}
\sy\!=\!\sum_{\si=1}^\sR \sgi \sri + \sv
$%\end{equation}
, where $\sgi$ denotes the complex coefficient of the ``forward'' channel between $\sRset_\si$ and $\sDset$, and $\sv\!\sim\!\sCN(0,\sNo)$ is complex Gaussian noise. 
Inserting \eqref{eq:xi} and \eqref{eq:asi} yields the compound channel model\footnote{Superscript $^T$ ($^H$) denotes (Hermitian) transposition; $\sdiag(x_1,\dots,x_m)$ is the $m\times m$ diagonal matrix with diagonal elements $x_1,\dots,x_m$.}
\begin{equation}
% \sy =  \svw^T \svhbar \ssymb + \svw^T \sGbar \svn +\sv = \shcomp\,\ssymb + \sucomp, \label{eq:y}
\sy =  \shcomp\ssymb + \sucomp, \quad\text{with }\,
\shcomp\define \svw^H \svhbar, \quad
\sucomp \define \svw^H \sGbar \svn +\sv.\label{eq:y}
\end{equation}
Here, $\svhbar \!\define\! [\shbar_1 \dots \shbar_\sR]^T$ with $\shbari\! \define\! \shi\sgi \spnorm \sqrt{\sPs}$, 
$\sGbar \!\define \!\sdiag(\sgbar_1,\dots,\sgbar_\sR)$ with  $\sgbari\! \define\! \sgi \spnorm $,
and $\svn\! \define\! [\sn_1 \dots \sn_\sR]^T$.
Since the weight vector $\svw\!\define\! [\sw_1 \dots \sw_\sR]^T$ enters also the noise part in \eqref{eq:y}, it demands careful design to prevent noise amplification.

From \eqref{eq:y}, the average power corresponding to the signal part of $\sy$ and the SNR at $\sDset$ 
% for given channel coefficients can be derived as %:$\sE\{|\svw^T\svhbar \ssymb|^2|\svh,\svg\}$$\sE\{|\sv+\svw^T\sGbar\svn|^2|\svh,\svg\}$
are respectively obtained as
\begin{align}
%\sPrx = \sE_{\ssymb,\svn,\sv}\{|\svw^T \svhbar \ssymb|^2\}= \svw^T \svhbar \svhbar^H \svw^\ast, \qquad \text{and} \label{eq:power}
\sPrx(\svw) & \,\define\, \sE\big\{|\shcomp \ssymb |^2 \big|\hspace*{.2mm} \svhbar\big\}\,= \,\svw^H \svhbar \svhbar^H \svw
\,=\, \big|\svw^H \svhbar\big|^2 
,
% \qquad\qquad \text{and} 
 \label{eq:power} \\[1.5mm]
%\end{equation}
%and
%\begin{equation}
%\sgamma = \frac{\sE_{\ssymb,\svn,\sv}\{|\svw^T \svhbar \ssymb|^2\}}{\sE_{\ssymb,\svn,\sv}\{|\svw^T \sGbar \svn +\sv|^2\}}=\frac{1}{\sNo}\frac{\svw^T \svhbar \svhbar^H \svw^\ast}{1+\svw^T\sGbar\sGbar^H\svw^\ast}, \label{eq:snr}
\ssnr(\svw) &\,\define\, \frac{\sE\big\{|\shcomp\ssymb|^2 \big|\hspace*{.2mm} \svhbar\big\}}{\sE\big\{|\sucomp|^2\big| \hspace*{.2mm}\sGbar\big\}}\,=\,\frac{1}{\sNo}\frac{\svw^H \svhbar \svhbar^H \svw}{1+\svw^H\sGbar\sGbar^H\svw} . \label{eq:snr}
\end{align}
% respectively. %Note that here expectation is only taken with respect to $\ssymb$, $\svn$, and $\sv$.
In the following, we will use $\sJ(\svw)$ as generic notation for our objective function, which can either be $\sPrx(\svw)$ or $\ssnr(\svw)$. The beamforming vector $\svw$ can be batch designed to maximize $\sJ(\svw)$ subject to a specific relay power constraint. %an per-relay or sum power constraint. 
We resort to two types of power constraints: % that can be imposed. 
% Since $\sE\{|\sxi|^2|\shi\} = \sPs |\shi|^2\!+\!\sNo$ ($\sE\{.\}$ denotes expectation), we write\footnote{The complex conjugation (superscript $^*$) of the beamforming weights is chosen to simplify notation later on.}
%\begin{equation} %\swi \frac{\sxi}{\sqrt{\sE\{|\sxi|^2\}}}=
%a_\si = \swi^* \gamma_\si \,,\quad\text{with }\,
%\gamma_\si \define \sqrt{\!\frac{\sP}{\sPs |\shi|^2 + \sNo}}. \label{eq:asi}
%\end{equation}
Constraining the complex beamforming weights to $|\swi|^2\!=\!1$ 
(this amounts essentially to phase-matching at the relays \cite{Mudumbai:2006aa}) ensures identical per-relay power $\sE\{|\sri|^2|\shi\}\!=\!\sP$.
In contrast, the total sum power constraint 
$\sE\big\{\sum_{\si=1}^{\sR}|\sri|^2|\sh_1,\dots,\sh_\sR\big\}\!=\!\sP$ requires that 
the beamforming vector has unit Euclidean norm, $\|\svw\|^2\!=\!1$.
However, such batch designs entail stringent requirements regarding the CSI available to the relays %(via direct estimation or feedback, 
(cf.\ Section \ref{sec:PA}).

% ************** PERTURBATION *********************************
\section{Adaptive Perturbation-based Beamforming}
\label{sec:perturb}
%%%%%%%%%%%%%%%%%%%%%%%%%%%%%%%%%%%%%%%%%%%%

\subsection{Transmission Principle}
%%%%%%%%%%%%%%%%%%%%%%%%%%%%%%%%%%%%%%%%%%%%
To avoid CSI at the relays, we study % next investigate in the following 
distributed beamforming using feedback-assisted adaptive weight perturbation. The idea underlying this approach is to maximize the objective function $\sJ(\svw)$ %a specific performance measure 
%at the destination 
%(signal power $\sPrx$ or SNR $\ssnr$) 
by adjusting the beamforming weights at the relays 
in an iterative manner using limited feedback (see Fig.\,\ref{fig:relay}).  
For co-located arrays with centralized processing similar ideas have been proposed in \cite{Banister:2003aa,Raghothaman:2003aa}.

%Transmission happens in frames consisting of a training sequence of length $\sTp$ and a data sequence of length $\sTd$.
Transmission happens in frames consisting of a training interval $\sTpset$ %of length $\sTp$ 
and a data interval $\sTdset$. %of length $\sTd$.
The relays use different beamforming weights to forward the training and data parts of each frame received from % $\sSset$ to the destination $\sDset$ 
$\sSset$ to $\sDset$ 
according to \eqref{eq:asi}. The idea is to apply the currently best beamforming
vector, denoted $\svwd$ ($\sk$ is the frame index), to the data while 
using a perturbed version $\svwt$ of the beamforming vector for the training portion. 
%Since $\sDset$ knows the training sequence, it can evaluate the effectiveness of the
%perturbed beamforming weights. It then provides the relays with one bit of feedback to indicate which beamforming vector shall be used to forward the data of the next frame.
%The effectiveness of the perturbed beamforming weights is then evaluated at $\sDset$. 
The destination evaluates the effectiveness of the perturbed weights and checks whether or not the perturbation improved the objective function $\sJ(\svwk)$. It then provides the relays with one bit of feedback to indicate which beamforming vector shall be used to forward the data of the next frame.

In the proposed scheme, the weights $\svwt$ %applied to the training signal
are obtained by an additive perturbation from the data beamforming vector $\svwd$. While in a centralized setup the perturbation vectors can be chosen randomly for each frame, our distributed setup necessitates a deterministic vector set, collected in a $\sR\times\sN$ matrix $\sQset=[\svq_0 \dots \svq_{N-1}]$, from which the perturbation vector is picked in a cyclic fashion (cf.\ Section \ref{ssec:pset}). 

In the following, we present two variants of the proposed PB-BF scheme and describe the individual steps in more detail.
For this discussion, we assume that all channel coefficients remain constant during the weight adaptation process.

\subsection{Take/Reject (T/R) Perturbation}
%%%%%%%%%%%%%%%%%%%%%%%%%%%%%%%%%%%%%%%%%%%%

 % for the $k$th frame 
The perturbed weights for the $\sk$th frame are computed as
\begin{equation}
\svwtu =
\svwd + \sbeta\, \svq_{\sk\,\text{mod}\,\sN} , \label{eq:TR}
%\svwhat =	\frac{\svwk + \sbeta\,\svq}{\|\svwk + \sbeta\, \svq \|},  \label{eq:TR}%&  \sfbmkp>\sfbmk, \\
\end{equation}
%\begin{equation}
%\svwhats \define 
%\begin{cases}
%\svw_\sk & \text{for}\;\sk=1,\\
%\svwk + \sbeta\, \svq & \text{for}\;\sk>1,
%\end{cases} \label{eq:TR}
%\end{equation}
where $\sbeta$ is a step-size parameter determining the adaptation rate, followed by
proper normalization, % to satisfy the power constraint, 
i.e.,
$\svwt\!=\!\svwtu/\|\svwtu\|$ in case of a sum power constraint
and $\swt\!=\!\swtu/|\swtu|$ for a per-relay power constraint.
% and normalized as $\svwhat\!=\!\frac{\svwhats}{\|\svwhats\|}$ 
The weights $\svwt$ are % kept constant during $\sTpset$ and 
applied to the training sequence received at the relays, which is then forwarded to $\sDset$. At $\sDset$, the known training sequence and the receive signal are used to evaluate the performance of $\svwt$ within $\sTpset$
according to the objective function, i.e., $\sfbmone\!=\!\sJ(\svwt)$. 
Recall that $\sJ(\cdot)$ represents either the received signal power $\sPrx(\cdot)$ in \eqref{eq:power} 
or the SNR $\ssnr(\cdot)$ in \eqref{eq:snr}. The actual estimation of these quantities will be addressed in 
Section \ref{sec:implementation}.

The destination then compares $\sfbmone$ to the performance $\sfbmtwo\!=\!\sJ(\svwd)$ achieved with the beamforming vectors $\svwd$ that 
up to this point performed best.
If $\sfbmone\le\sfbmtwo$, $\svwt$ does not perform better than $\svwd$ and hence the relays should stick with $\svwd$ for the
data in the next transmission frame (``reject'' $\svwt$). Otherwise ($\sfbmone>\sfbmtwo$), the beamforming vector $\svwt$ improves on $\svwd$
and should thus be used in the next frame to transmit the data (``take'' $\svwt$). This rationale can be implemented be letting $\sDset$ provide the relays
with a single bit of feedback, given by
\begin{equation*}
\sfbk = \su(\sfbmone \!-\! \sfbmtwo), %\label{eq:FB}
\end{equation*}
where $\su(.)$ denotes the unit step function.
Depending on the feedback bit, the relays update the data beamforming vector for the next frame as
\begin{equation*}
\svwkp =
\begin{cases}
\svwd, &  \text{if }\sfb_\sk = 0, \\ 
\svwt, &  \text{if }\sfb_\sk = 1. 
\end{cases}
% \frac{\svwk + \sfbup(\sfb)\, \sbeta\, \svq}{\|\svwk + \sfbup(\sfb)\,\sbeta\,\svq \|},  
%\label{eq:update}
%\svwkps = \svwk + \sbeta\,\sfb(\sfbm)\, \svq,
\end{equation*}
The vector $\svwkp$ will be the basis for the next perturbation according to \eqref{eq:TR}. The destination $\sDset$ performs the corresponding update $\sfbmkp\!=\!\max\{\sfbmone,\sfbmtwo\}$. This process continues in an iterative manner. 
During the first frame, the scheme is initialized %by setting all elements of $\svw_0$ and $\tilde\svw_0$ equal.
with $\tilde\svw_0\!=\!\svw_0$ (by setting, e.g., $\svw_0\!=\![1 \dots 1]^T\!/\!\sqrt{\sR}$ in case of a sum power constraint) 
and $\sfbm_0\!=\!0$. 
% In the first frame ($\sk\!=\!1$) the weight vector $\svwhat$ is initialized with unperturbed weights in order to evaluate a reference measure $\sfbmtwo$ at $\sDset$. For all frames with $\sk\!>\!1$, $\sDset$ updates $\sfbmtwo\!=\!\sfbmone$, if $\sfbmone\!>\!\sfbmtwo$, i.e., if the perturbation improves the performance at $\sDset$.

\subsection{Plus/Minus (P/M) Perturbation}
T/R perturbation has the advantage that performance never deteriorates, i.e., $\sJ(\svwkp)\!\ge\!\sJ(\svwk)$.
On the other hand, in many cases the perturbation \eqref{eq:TR} will not yield an improvement, which entails $\svwkp\!=\!\svwd$ and hence slow adaptation. We next discuss an alternative perturbation scheme with faster adaptation rate. 
Here, the training interval $\sTpset$ is split into two halves $\sTone$ and $\sTtwo$  
for which different perturbed beamforming vectors are used, 
% to these two halves, 
i.e.,\footnote{These expressions are valid for the sum power constraint. With a per-relay power constraint, weight normalization has to be performed element-wise.}% normalization of the weights is required.}%weight normalization has to be performed element-wise.} 
% equal time intervals $\sTone$ and $\sTtwo$. 
\begin{equation}
%\svwhats 
\svwta \!= \!
% \frac{\svwd + \sbeta\, \svq_{(k\text{ mod}\,N)} {\|\svwk + \sbeta \svq \|}   ,\text{within} \;\sT_1, \\
\frac{ \svwk + \sbeta\, \svq_{ \sk\,\text{mod}\,\sN } }{\|\svwk \!+\! \sbeta\, \svq_{\sk\,\text{mod}\,\sN}\|},
% & \text{for} \;\sTone, 
% \\
%\frac\svwk - \sbeta \,\svq{\|\svwk - \sbeta \svq \|}& \text{within} \;\sT_2.
\quad \svwtb \! = \!
\frac{ \svwk - \sbeta\, \svq_{\sk\,\text{mod}\,\sN} }{\|\svwk \!-\! \sbeta\, \svq_{\sk\,\text{mod}\,\sN}\|} .
% \svwk - \sbeta \,\svq & \text{for} \;\sTtwo,
\label{eq:PN}
\end{equation}
% and normalized as $\svwhat\!=\!\frac{\svwhats}{\|\svwhats\|}$. 
The destination $\sDset$ then measures the performance of $\svwta$ and $\svwtb$
by evaluating the objective function according to
$\sfbmonep\!=\!\sJ(\svwta)$ and $\sfbmonem\!=\!\sJ(\svwtb)$ within $\sTone$ and $\sTtwo$, respectively.
While in principle we could pick the beamforming weights corresponding to the maximum of $\sfbmonep$, $\sfbmonem$, and $\sfbmtwo\!=\!\sJ(\svwd)$ (the performance of the current data beamforming vector), 1-bit feedback can only support binary choices. Hence, $\svwd$ will be discarded in any case. $\sDset$ broadcasts the feedback bit
$\sfbk = \su(\sfbmonem \!\!-\hspace*{-.2mm}  \sfbmonep)$ to the relays, indicating whether the 
``plus'' perturbation $\svwta$ or the ``minus'' perturbation $\svwtb$ performs better.
In the next frame, the relays use the beamforming vector
\begin{equation*}
\svwkp =
\begin{cases}
\svwta, &  \text{if }\sfbk = 0, \\ 
\svwtb, &  \text{if }\sfbk = 1. 
\end{cases}
% \frac{\svwk + \sfbup(\sfb)\, \sbeta\, \svq}{\|\svwk + \sfbup(\sfb)\,\sbeta\,\svq \|},  
%\label{eq:update2}
%\svwkps = \svwk + \sbeta\,\sfb(\sfbm)\, \svq,
\end{equation*}
%While P/M perturbation typically converges faster than T/R, sometimes both perturbations in \eqref{eq:PN} will
%deteriorate performance, and hence P/M performance never totally saturates. Furthermore, 
%only half of the training interval can be used to estimate each of $\sfbmonep$ and $\sfbmonem$.
Although P/M perturbation shows typically faster adaptation than T/R, sometimes both perturbations in \eqref{eq:PN} deteriorate the performance with respect to $\sfbmtwo$.
%deteriorate performance
Thus, P/M performance may fluctuate continually. %never totally saturates. 
Furthermore, only half of the training interval can be used to estimate each of $\sfbmonep$ and $\sfbmonem$.

\subsection{Perturbation Set}
\label{ssec:pset}
Vector normalization of the weights in \eqref{eq:TR} and \eqref{eq:PN} ensures that 
the sum power constraint is satisfied, but 
% guarantees that $\|\svw\|^2\!=\!1$, thus implying an aggregate relay power constraint $\sPsum\!=\!\sPr$ and a per-relay power $\sE\{|\sri|^2\}\!\le\!\sPr$. 
requires that each relay knows all elements of the beamforming vector. 
% Exchanging all weights between the relays would imply a tremendous signaling overhead which is quite unrealistic. 
%Hence, the {\em stochastic} gradient algorithm with random perturbation vectors 
%that is used in the centralized case \cite{Banister:2003aa} cannot be applied to relay networks.
%% which makes use of a {\em stochastic} gradient algorithm that randomly generates a perturbation vector with i.i.d. complex Gaussian entries $\sCN(0,1)$. 
%% In order to avoid weight sharing, 
%Rather, we propose to use a matrix $\sQset$ of {\em deterministic} perturbation vectors (cf.\ \cite{Raghothaman:2003aa}) known to each relay. 
%This allows each relay to keep track of the full beamforming vectors and hence to perform vector normalization.
Hence, a {\em stochastic} gradient algorithm with random perturbation vectors 
(as in \cite{Banister:2003aa}) cannot be applied to relay networks; this would require to exchange all weights among the relays, thus imposing a tremendous signaling overhead.
% which makes use of a {\em stochastic} gradient algorithm that randomly generates a perturbation vector with i.i.d. complex Gaussian entries $\sCN(0,1)$. 
% In order to avoid weight sharing, 
Rather, we propose to use a matrix $\sQset$ of {\em deterministic} perturbation vectors (cf.\ \cite{Raghothaman:2003aa}) known to each relay. 
This allows each relay to keep track of {\em all} beamforming weights and to perform vector normalization locally.
% Hence, each relay can independently update/calculate also the weights of the other relays participating in the transmission. 
%We note that $\sQset$ should be chosen such that $\svwk$ is uniformly distributed on the $\sR$-dimensional complex hypersphere. 
Reasonable choices for the deterministic perturbation matrix are
$\sQset\!=\![\sQ,j\sQ]$ (i.e., $\sN\!=\!2\sR$ vectors) for P/M perturbation and $\sQset\!=\![\sQ,j\sQ,-\sQ,-j\sQ]$ ($\sN\!=\!4\sR$) for T/R perturbation; here, $\sQ$ is an $\sR\!\times\!\sR$ unitary matrix. %, 
%e.g., the discrete Fourier transform (DFT) matrix. 
We observed that choosing $\sQ$ as discrete Fourier transform (DFT) matrix provides good performance under various conditions. %and convergence behavior
% The perturbation vector $\svq$ is then selected in a pre-determined order from $\sQset$. 
% We note that the perturbation/update complexity increases linearly with increasing number of relays. 

Note that with the per-relay power constraint, element-wise normalization does not require knowledge of all weights at each relay, %in T/R and P/M perturbation is done for each relay separately
thus allowing also for stochastic perturbations.

% If no power control is required and the relays are allowed to transmit with their highest power $\sPr$, the vector normalization in \eqref{eq:TR}, \eqref{eq:PN}, and \eqref{eq:update} can be replaced by component-wise normalization given by $\swi\!=\!\frac{\swis}{|\swis|}$. In this case the aggregate relay power equals $\sPsum\!=\!\sPr\sR$ which is $\sR$-times higher than for vector normalization. Knowledge of all weights at each relay is then no longer required and hence a stochastic perturbation approach can be used. Here, the update and perturbation complexity does not depend on the number of relays $\sR$.

% \section{Implementation Aspects}

\subsection{Channel, Power, and SNR Estimation} 
\label{sec:implementation}

We next discuss the estimation of the receive signal power $\sPrx(\svw)$ in \eqref{eq:power} and the SNR $\ssnr(\svw)$ in \eqref{eq:snr} which are used as performance measures, as well as the estimation of the compound channel $\shcomp$ in \eqref{eq:y} required for coherent detection.
% at the destination. 
%$\sPs$ equal for pilots and data

%We introduce the time index $\st$.
%\vspace*{1mm}
In the following, we omit the frame index $\sk$ and denote the pilot sequence within a transmission frame as
$\spilot[\st]$, $\st\!\in\!\sTsets$.
% is a (auxiliary) set of training time slots of length $\sTps$ corresponding to a specific transmission frame. 
%Moreover, we assume equal average pilot and data signal power $\sPs\!=\!\sE\{|\spilot[n]|^2\}$.
% Assuming that the pilots are known at $\sDset$, a 
The destination can then compute the maximum likelihood (ML) estimate of the compound channel as
\begin{equation}
\shcomphat = \frac{\sum_{\st\in\sTsets}\sy[\st]\spilot^\ast[\st]}{\sum_{\st\in\sTsets}|\spilot[\st]|^2}. \label{eq:hest}
%\frac{1}{\sTps}\sum_{\st\in\sTsets} \frac{\sy[\st]}{\spilot[\st]}. \label{eq:hest}
\end{equation}
Using \eqref{eq:hest}, the ML estimates of receive signal power and SNR can be obtained as
\begin{equation}
\sPrxhat = |\shcomphat|^2, \qquad
%|\frac{\sum_{\st\in\sTsets}\sy[\st]\spilot^\ast[\st]}{\sum_{\st\in\sTsets}|\spilot[\st]|^2}|^2 
\ssnrhat = \frac{|\shcomphat|^2}{\frac{1}{|\sTsets|}\sum_{\st\in\sTsets}\bigl|\sy[\st]-\shcomphat\spilot[\st]\bigr|^2}.
\label{eq:snrest}
% \label{eq:Prxest}
%\biggl|\frac{1}{\sTps}\sum_{\st\in\sTsets} \frac{\sy[\st]}{\spilot[\st]}\biggr|^2. \label{eq:Prxest}
\end{equation}
% Naturally, the estimates improve with increasing training length $\sTps$.

For T/R perturbation, \eqref{eq:hest} and \eqref{eq:snrest} are evaluated using
$\sTsets\!=\!\sTpset$. %The destination stores the channel estimate and uses it for detection until a weight update occurs, in which case it switches to the new estimate $\shcomphat$ obtained during the previous training period. 
After each weight update the destination stores the channel estimate and uses it for data detection of the subsequent frames till the next update occurs. %In this case it switches to the new estimate $\shcomphat$ obtained during the previous training period. 
%
% To obtain the performance measure $\sfbmone$ and the reference measure $\sfbmtwo$ used by the {\em T/R-perturbation} technique, we first compute the estimate of the corresponding compound channel by averaging over the entire training interval $\sTpset$ in \eqref{eq:hest}, and then we insert this estimate in \eqref{eq:Prxest} or \eqref{eq:snrest}, respectively (setting $\sTsets\!=\!\sTpset$ with $\sTps\!=\!\sTp$). %The same settings are used to compute the compound channel estimate in \eqref{eq:hest}. 
% The detector at $\sDset$ uses the compound channel estimate which has been evaluated together with the reference measure $\sfbmtwo$. This estimate is then updated, whenever $\sfbmone\!>\!\sfbmtwo$. More frequent updates have to be forced explicitly. %of the channel estimate at $\sDset$ 
%
With P/M, \eqref{eq:hest} and \eqref{eq:snrest} are calculated twice in each frame with $\sTsets\!=\!\sTone$ and $\sTsets\!=\!\sTtwo$. 
The channel estimate corresponding to the better beamforming vector is then kept for data detection in the next frame.
Alternatively, an approximate ML estimate for the channel coefficient $\shcomp$ can be obtained by evaluating \eqref{eq:hest} over the whole training interval ($\sTsets\!=\!\sTpset$) within the same frame (cf.\,\cite{Banister:2003aa}), provided that the step-size $\sbeta$ is chosen sufficiently small and $|\sTone|\!=\!|\sTtwo|$. 

% In a similar way the performance measures can be computed for the {\em P/M perturbation} case. Here, $\sfbmone$ is evaluated over the training interval $\sTone$ with length $\sTpone$ (corresponding to the ``plus'' perturbation) using \eqref{eq:hest}, \eqref{eq:Prxest}, and \eqref{eq:snrest}, whereas $\sfbmtwo$ is obtained by averaging over $\sTtwo$ with $\sTptwo$ (corresponding to the ``minus'' perturbation). If $\sTpone\!=\!\sTptwo$, the compound channel used for detection at $\sDset$ can be simply estimated by averaging over the entire training interval, i.e., using $\sTsets\!=\!\sTpset$ in \eqref{eq:hest} (cf.\,\cite{Banister:2003aa}). 

\subsection{Birth and Death of Relays}
\label{sec:implementation2}

Our deterministic perturbation approach is scalable in that it can be easily adapted to deal with the situation where relays enter (``birth'') or leave (``death'') the network, even in the case of a sum power constraint.
% , whereas the relays are not necessarily able to share information within each other. %Therefore, the event of birth or death is treated via $\sDset$. 
We assume that the maximum number of relays is $\sRmax$, of which $\sR\!\le\!\sRmax$ are active and can exchange information with $\sDset$ but not with each other. 
%All relays know their ``identity'' (index $i$), which is fixed and allows them to pick their corresponding beamforming/perturbation weight. % and perform vector normalization.
%All relays know their ``identity'' (index $i$),
%which is fixed and allows them to pick the corresponding perturbation weight in order to compute their own beamforming weight. % and perform vector normalization.
%In essence, the destination and the relays then only need to keep track of the active relays. 
In essence, the destination and each relay keep track of all the active relays. The relays can then compute the required vector norm locally. % such tto compute the vector norm locally. 
Additionally, all relays know their ``identity'' (index $i$), which is fixed and enables them to pick their corresponding beamforming weight. % in order to compute their own beamforming weight. % and perform vector normalization.
If a relay $\sRset_{i_0}$ drops out, it informs $\sDset$ which in turn broadcasts the relay index $i_0$ to the remaining relays using  $\log_2(\sRmax)$ bits. These relays then exclude the corresponding beamforming/perturbation weight from the update process. 
If a new relay enters the system, it contacts $\sDset$ which in turn broadcasts 
$\sRmax$ bits to indicate to all relays (also to the new one) which relays are active. 
Since the new relay cannot know the current beamforming weights of the other relays, 
the weight adaptation process needs to be re-initialized in this case.
% started all weights we propose to re-initialize all weights and restart the weight adaptation process.

In the case of a per-relay power constraint, element-wise weight normalization allows that the relays only need to track their own weights. This renders a birth-and-death protocol particularly easy, since relays can enter or leave the system completely arbitrarily without informing the other relays.

% ************** POWER ALLOCATION *********************************
\section{Comparison with Optimal Beamforming} % and Power Allocation Schemes}
\label{sec:PA}
We next compare optimal batch beamforming designs with adaptive PB-BF. The former % (i.e., with power allocation) 
% power control methods requiring CSI at the relays 
%with . The main difference lies in the fact that for the batch designs 
requires each relay having either local CSI (i.e., each relay's own back- and forward channel) or global CSI (i.e., all channels) available, whereas PB-BF exploits limited feedback to avoid CSI at the relays. 

\subsection{Optimal Batch Designs}
{\em Equal Gain Combining (EGC).} %If the cost function equals the 
Maximizing $\sPrx(\svw)$ or $\ssnr(\svw)$ under a per-relay power constraint
yields the beamforming weights 
%\begin{equation} 
% $\swi^{(\text{EGC})}=
$\swi =
{\shbari}/{|\shbari|}=
{\shi\sgi}/{|\shi\sgi|}$ % \label{eq:egc-wopt}
%\end{equation}
that amount to coherent combining \cite{dana03}. This scheme requires that each relay knows the phase of its backward and forward channel. 
% yielding the average receive signal power  
% \begin{equation}
% \sPrx^{(\text{EGC})} = \left|\sum\nolimits_{\si=1}^\sR |\shbari|\right|^2 \label{eq:egc-power}
% \end{equation}
% at $\sDset$.
% It can be seen that the signal part of the received signal $\sy$ in \eqref{eq:y} adds up coherently, but the noise part does not. %Since $|\swi|\!=\!1$ across $\si$, the relays cooperate with equal (maximum) transmit power $\sPr$, resulting in the average sum power $\sPsum=\sPr\sR$. 

% Alternatively, the receive power in \eqref{eq:egc-power} can be approached using feedback-assisted T/R- (cf.\;\cite{Mudumbai:2006aa,Thukral:2007ab}) or P/M perturbation techniques and a power-based performance measure in \eqref{eq:Prxest}. Thereby, component-wise weight normalization fulfills the per-relay power constraint implied by $\left|\swi^{(\text{EGC})}\right|\!=\!1$.
\vspace*{1mm}
{\em Power Maximization under Sum Power Constraint (P-SP).}
Optimizing $\sPrx(\svw)$ in \eqref{eq:power} under a sum power constraint amounts to maximizing $|\svw^H\svhbar|$ subject to $\|\svw\|^2\!=\!1$.
Via the Cauchy-Schwarz inequality, the solution is obtained as 
% represents a generalized eigenvalue problem which can be easily solved due to the fact that $\svhbar\svhbar^H$ in \eqref{eq:power} constitutes a rank-1 matrix. The optimum beamforming vector is given by 
% $\sw_i^{(\text{PSP})} 
$\swi\!=\!{\shbari}/{\|\svhbar\|}$
requiring global CSI at $\sRset_\si$. Alternatively, if global CSI is available at $\sDset$, each relay needs only local CSI and feedback of $\|\svhbar\|$ from $\sDset$. 
% $\svw^{(\text{PSP})}\!\define\!\frac{\svhbar}{\|\svhbar\|}$. 
%\begin{equation}\sPrx^{(\text{PSP})} = \|\svhbar\|^2. \label{eq:psp-power}\end{equation}
%Assuming a sum relay power constraint $\|\svw\|^2\!=\!1$ on the weights. %Note that such a constraint still respects the per-relay power constraint $\sE\{|\sri|^2\}\!\le\!\sPr$, however, 
% Note that here the aggregate relay power $\sPsum$ is $\sR$-times smaller than for EGC.  %The solution to this generalized eigenvector problem is given by the principal eigenvector of $\svhbar\svhbar^H$ which constitutes a rank-1 matrix, and thus can be written as %the optimal beamforming vector solving this problem is given by
%\begin{equation}
%\svw^{(\text{PSP})}=\frac{\svhbar^\ast}{\|\svhbar\|}. \label{eq:psp-wopt}
%\end{equation}
This shows that the relays optimally allocate their transmit power to match the current local fading coefficients while performing coherent combining.
% or local CSI and feedback in form of $\|\svhbar\|$ from $\sDset$. 
% (i.e, relays which experience strong fading, forward their signal with less power than others). %Inserting \eqref{eq:psp-wopt} in \eqref{eq:power} then results in 
%\begin{equation}
%\sPrx^{(\text{PSP})} = \|\svhbar\|^2. \label{eq:psp-power}
%\end{equation}
% Comparing \eqref{eq:egc-power} and \eqref{eq:psp-power}, it can be seen that $\sPrx^{(\text{PSP})}\!<\!\sPrx^{(\text{EGC})}$.

\vspace*{1mm}
{\em SNR Maximization under Sum Power Constraint (S-SP).}
% The PSP approach can be modified such that the 
The beamforming vector $\svw$ can also be chosen to maximize the 
SNR $\ssnr(\svw)$ in \eqref{eq:snr} 
% represents the cost function to be maximized 
\cite{Larsson:2003aa,Hammerstrom:2004aa}. % subject to the sum power constraint $\sPsum\!=\!\sPr$ . Following the constraint $\|\svw\|^2\!=\!1$, \eqref{eq:snr} can be 
Under the sum power constraint $\|\svw\|^2\!=\!1$ this can be shown to lead to a generalized 
eigenvalue problem whose solution is ($\seye$ is the identity matrix) \cite{Hammerstrom:2004aa}
\begin{align}
\svw &= \frac{(\seye+\sGbar \sGbar^H)^{-1}\svhbar}{\|(\seye+\sGbar \sGbar^H)^{-1}\svhbar\|}. \label{eq:larsson-wopt}
\end{align}
%
% Reformulating \eqref{eq:snr} as ($\seye_m$ denotes a $m\times m$ identity matrix)
% \begin{equation*}\ssnr = \frac{\svw^T\svhbar\svhbar^H\svw^\ast}{\svw^T(\seye_\sR+\sGbar \sGbar^H)\svw^\ast} %\label{eq:larsson-snr}\end{equation*}
% with $\|\svw\|^2\!=\!1$, 
%Maximization of \eqref{eq:larsson-snr} over $\svw$ with constraint $\|\svw\|^2\!=\!1$ 
%again forms a general eigenvector problem which can be easily solved due to the fact that the matrix ($\seye_\sR+\sGbar \sGbar^H$) is a positive semidefinite, diagonal matrix. %and $\svhbar\svhbar^H$ being a rank-1 matrix. 
%The optimum solution is then given by
%\begin{align}
%\svw^{(\text{SSP})} &= \frac{(\seye_\sR+\sGbar \sGbar^H)^{-1}\svhbar^\ast}{\|(\seye_\sR+\sGbar \sGbar^H)^{-1}\svhbar\|} \qquad \text{with} \label{eq:larsson-wopt}\\
%\end{equation}
%with global extremum
%\begin{equation}
%\sgamma^{(\text{SSP})} &= \frac{1}{\sNo}\svhbar^H(\seye_\sR+\sGbar \sGbar^H)^{-1}\svhbar. \label{eq:ssp}
%\end{align}
Again this essentially requires either
global CSI at the relays or local CSI with feedback of $\|(\seye+\sGbar \sGbar^H)^{-1}\svhbar\|$ from $\sDset$.
%Inserting \eqref{eq:larsson-wopt} in \eqref{eq:larsson-snr} yields the corresponding receive SNR 
In contrast to P-SP, \eqref{eq:larsson-wopt} also accounts for noise amplification.
% , and thus this method achieves optimal relay power control under the assumed sum power constraint.

%\subsection{SNR-optimal BF under Individual Power Constraints (SIP)}
%Alternatively, the scheme in \cite{Jing:2007aa}\footnote{Therein denoted as {\em network beamforming}.} maximizes \eqref{eq:snr} under the more practical assumption of individual power constraints $\sE\{|\sri|^2\}\!\le\!\sP_{\sr_\si}\!\le\!\sPr$ at the relays. Although the scheme does not consider a sum power constraint, the aggregate power is naturally bounded as $\sPsum\!\le\!\sPr\sR$. Due to space reasons, we do not review the computation of the beamforming weights in detail here (see \cite{Jing:2007aa}). Nevertheless, we include this scheme for comparison in the numerical analysis in Section \ref{sec:results}. Calculation of the weights requires global CSI at the relays, however, distributive feedback strategies can be found in \cite{Jing:2007aa}.

\subsection{Comparison with PB-BF Schemes}
It can be shown that %the respective objective function $\sJ(\svw)$ 
$\ssnr(\svw)$ and $\sPrx(\svw)$ have 
%local maxima% under a per-relay or a sum power constraint
%, and the 
only a global maximum (unique up to phase ambiguity) under both power constraints
and this maximum is achieved by the corresponding optimal batch design. %(presented in the previous section). 
The proposed PB-BF schemes aim to maximize $\ssnr(\svw)$ or $\sPrx(\svw)$, and indeed approach their optimal counterparts (cf.\ Section \ref{sec:results}). EGC can be approximated by PB-BF using element-wise normalization and the objective function $\sJ(\svw)$ chosen as received signal power $\sPrx(\svw)$. 
%The associated PB-BF scheme that is able to approach the performance of P-SP and S-SP uses $\sJ(\svw)\!=\!\sPrx(\svw)$ and $\sJ(\svw)\!=\!\ssnr(\svw)$ as objective function, respectively, and vector normalization of the beamforming weights.
P-SP and S-SP performance can be approached using $\sJ(\svw)\!=\!\sPrx(\svw)$ and $\sJ(\svw)\!=\!\ssnr(\svw)$ as objective function, respectively, and vector normalization of the beamforming weights.
%S-SP performance can be approximately obtained by using %The corresponding adaptive PB-BF scheme that allows to approach % \eqref{eq:ssp} employs a SNR-based performance measure (e.g.\;\eqref{eq:snrest}), deterministic perturbation set, 
% and performing vector normalization. 
The PB-BF schemes can be implemented via T/R or P/M perturbation.

\begin{figure*}[t]
\psfrag{D}{\scriptsize\hspace*{-.1mm}$\bar{\text{\sf P}}$}
\centering
\subfigure[]{\includegraphics[scale=0.31]{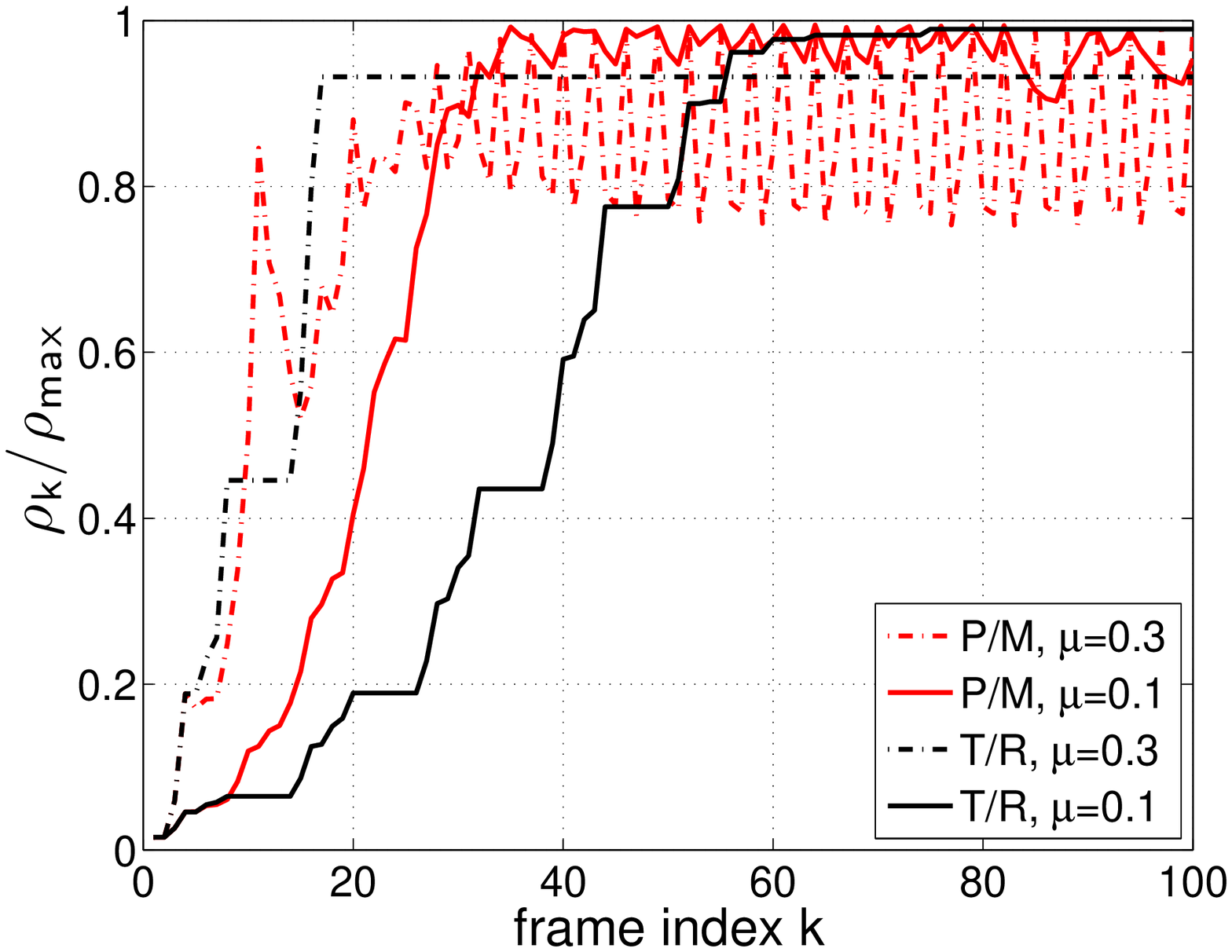}}
\hfill
\subfigure[]{\includegraphics[scale=0.31]{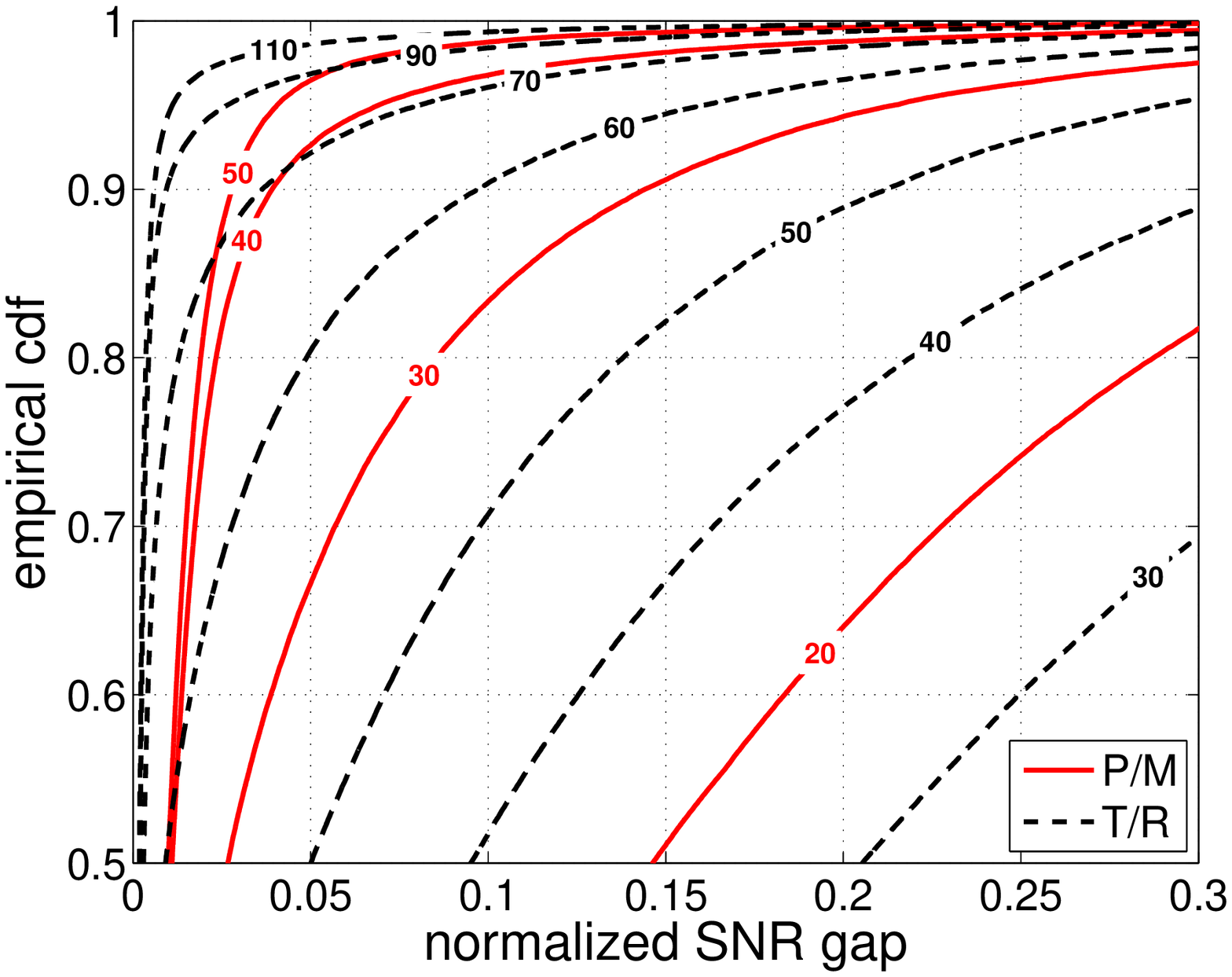}}
\hfill
\subfigure[]{\includegraphics[scale=0.31]{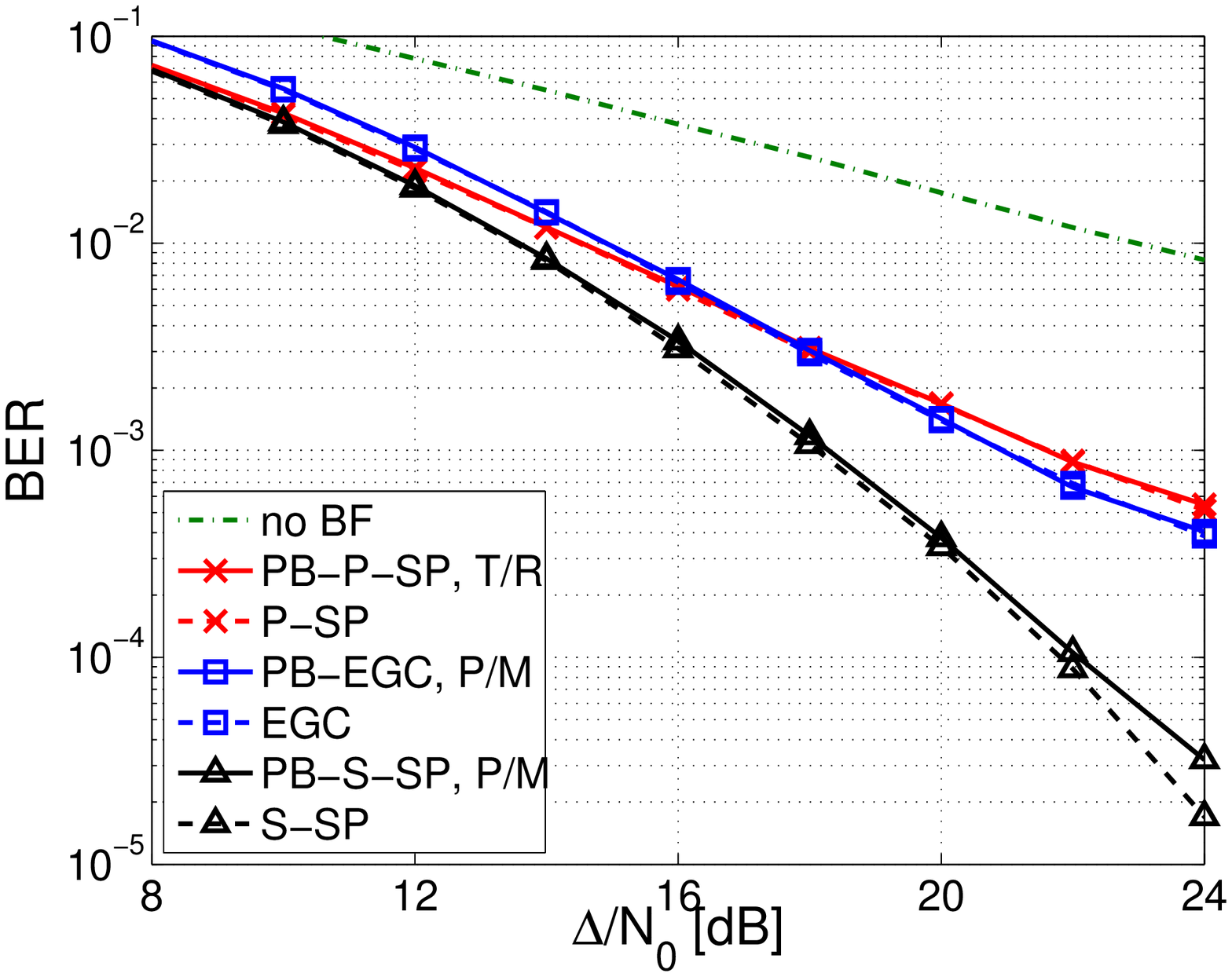}}
%\subfigure[]{\includegraphics[scale=0.38]{plots/plot_cap_4x2_16QAM_all.eps}}
%\vspace*{-2mm}
\vspace*{-1mm}
\caption{PB-BF performance for idealized scenario with static channels in a 3-relay network: (a) example for SNR evolution at $\sDset$ ($\sPsum/\sNo=18\,$dB), (b) cdf of the SNR gap after a fixed number of frames ($\sPsum/\sNo=18\,$dB, $\sbeta=0.1$), and (c) BER versus nominal SNR ($\sbeta=0.1$).}
\label{fig:ideal}
\vspace*{-2mm}
\end{figure*}
%\begin{figure}[ht]
%\centering
%\includegraphics[scale=0.33]{plots/plot_conv_R2.eps}
%%\hfill
%%\subfigure[]{\includegraphics[scale=0.38]{plots/plot_cap_4x4_16QAM_all.eps}}
%%\hfill
%%\subfigure[]{\includegraphics[scale=0.38]{plots/plot_cap_4x2_16QAM_all.eps}}
%%\vspace*{-2mm}
%\caption{Convergence for static channels with $\sP\!=\!12$\,dB and $\sbeta\!=\!0.1$.}
%\label{fig:conv}
%%\vspace*{-2mm}
%\end{figure}
%
%\begin{figure}[ht]
%\centering
%\includegraphics[scale=0.33]{plots/plot_hist_break.eps}
%\caption{Convergence rate.}
%\label{fig:conv-rate}
%%\vspace*{-2mm}
%\end{figure}
%\begin{figure}[ht]
%\centering
%\includegraphics[scale=0.33]{plots/plot_ber_R2.eps}
%\caption{Performance comparison of perturbation-based BF versus optimal BF schemes under Rayleigh fading.}
%\label{fig:ber}
%%\vspace*{-2mm}
%\end{figure}

% ************** SIMULATION RESULTS *********************************
\section{Simulation Results}
\label{sec:results}
We next investigate a network with $\sR=3$ relays via numerical simulations; 
we will refer to the PB-BF schemes by adding the prefix `PB-' to the corresponding batch design.
% and compare the optimal beamforming schemes from Section \ref{sec:PA} with their perturbation-based pendants. %for serve as benchmarks. 
%, negligible measurement/feedback latency,
For a fair comparison, all schemes use the same total relay power $\sPsum$ (in \eqref{eq:asi} we thus have $\sP\!=\!\sPsum/\sR$ under a per-relay power constraint and $\sP\!=\!\sPsum$ under a sum power constraint). The source $\sSset$ transmits BPSK symbols with transmit power $\sPs\!=\!\sPsum$ and the destination $\sDset$ employs an ML detector. 
%??% However, the resulting aggregate relay transmit power of the various schemes might not be equal. 
We further assume error- and delay-free 1-bit feedback, and employ a deterministic perturbation set based on a $3\times 3$ DFT matrix. %If not denoted otherwise, $\sbeta\!=\!0.1$ and $\sP\!=\!12$\,dB. %,, can be evaluated directly.   

\subsection{Idealized Scenario}
\label{ssec:ideal}
In this scenario, all channels are static i.i.d.\ Rayleigh fading with different path loss, i.e., $\shi,\sgi\!\sim\!\sCN(0,d_\si^{-2})$ with $d_\si\!=\!1,3,5$. %, but remain static within each frame. 
% channels (without path-loss effects). 
Each relay perfectly knows its backward channel (used in \eqref{eq:asi}), and $\sDset$ has perfect knowledge of the compound channel $\shcomp$ and the  performance measures $\sPrx(\svw)$ and $\ssnr(\svw)$. % are evaluated exactly at $\sDset$. 
Unless stated otherwise, %the adaption parameter 
% $\sbeta=0.1$ and 
$\sPsum/\sNo\!=\!18$\,dB.  

\vspace*{1mm}
{\em Convergence Behavior.}
% Assuming static channels, 
For the case of PB-S-SP using P/M  
and T/R perturbation with step size $\sbeta\!=\!0.1$ and $\sbeta\!=\!0.3$,
Fig.\ \ref{fig:ideal}(a) shows the evolution of the receive SNR $\ssnrk\!=\!\ssnr(\svwk)$ (normalized by the maximum receive SNR) versus the frame index $k$ for one channel realization.
% In particular, we plot the ratio between the receive SNR of the $\sk$th update, denoted by $\sgamma_\sk$, and the optimal SNR value in \eqref{eq:ssp}, denoted by $\sgamma_\text{opt}$. 
It is seen that with T/R $\ssnrk$ is nondecreasing and reaches almost optimal performance;
a larger step size results in faster convergence but also in a larger gap to the optimum.
Similar observations apply to P/M, which converges significantly faster than T/R, but features continual fluctuations whose amplitude increases with the step size.
%whereas P/M eventually leads to fluctuations whose amplitude 
%is larger for larger step size. 

% shows fluctuations when using P/M perturbation. The adaption parameter $\sbeta$ influences the convergence rate such that larger $\sbeta$ allows faster convergence. However, larger $\sbeta$ also implies larger fluctuations in the P/M case, and may prevent $\sgamma_\sk$ from converging close to the optimum in the T/R-case. 
%\vspace*{1mm}
%\noindent{\bf Convergence Rate.} 

For a systematic assessment of the convergence rate of PB-S-SP (with $\sbeta\!=\!0.1$),
Fig.\ \ref{fig:ideal}(b) shows %provides a systematic assessment of the convergence rate for PB-S-SP more systematically by plotting 
the empirical cumulative distribution function (cdf) of the normalized SNR gap
that remains after a certain number of frames
% updates 
(shown as curve labels). The cdfs were obtained with $10^5$ fading realizations.
% It can be seen that 
P/M converges considerably faster than T/R. 
To achieve an SNR gap of less than $4.3\%$ in $91\%$ of the cases,
P/M and T/R respectively require 40 and 70 iterations. 
However, after a large number of frames, T/R on average features a considerably smaller SNR gap than P/M.
Our simulations also revealed that a larger number of relays leads to slower convergence; for space reasons, the corresponding curves cannot be shown here.

%\subsection{BER Results}
\vspace*{1mm}
{\em BER Performance.}
Fig.\,\ref{fig:ideal}(c) plots bit-error rate (BER) versus 
nominal SNR $\sPsum/\sNo$ (in dB) for the batch designed beamforming schemes EGC, P-SP, S-SP, and their
perturbation-based counterparts. % transmit power $\sP$ (in dB). 
In each simulation run, only the frames after convergence of the PB-BF schemes were taken into account for the BER evaluation (again $\sbeta\!=\!0.1$).
% For each fading realization we perform a training phase of $60$ updates such that the weights are able to ``learn'' the channels before data transmission. %The transmission for each fading block was split into a training phase with $60$ updates and a data phase in order to guarantee convergence before data transmission. 
%We compare the optimal BF schemes in Section \ref{sec:PA} with their perturbation-based pendants. 
As a reference, we include an AF scheme that uses %$\sw_1\!=\!\sw_2\!=\!1/\sqrt{3}$, i.e., 
uniform PA and no coherent combining % does not use beamforming and employs equal relay transmit power 
(labeled `no BF'). %Note that in contrast to the other schemes, EGC and no-BF employ a relay sum power of $\sR \sP$ instead of $\sP$.
%, as well as the beamforming scheme presented in \cite{Jing:2007aa} (`SIP'); here, the optimal beamforming weights are obtained by maximizing the SNR in \eqref{eq:snr} under the assumption of individual power constraints $\sE\{|\sri|^2\}\!\le\!\sP_{\sr_\si}\!\le\!\sPr$ at $\sRset$. Although this scheme does not consider a sum power constraint, the aggregate relay power is naturally bounded as $\sPsum\!\le\!\sPr\sR$.

It can be seen that all PB-BF performance curves are
almost indistinguishable from those of their corresponding batch designs  % optimal schemes. %A significant performance 
and offer significant gains over the no-BF case (e.g., 8\,dB SNR improvement at a BER of $10^{-2}$).
SNR optimization (PB-S-SP) is seen to outperform power optimization (PB-P-SP) at high SNR.
In fact, PB-S-SP and S-SP are the only schemes to achieve a diversity larger than 1. %order of 3. 
Power optimization under a sum power constraint
%Non-uniform PA (performed by 
(PB-P-SP) and %uniform PA 
under a per-relay power constraint (PB-EGC) perform almost identically; in fact, PB-P-SP 
appears to suffer from noise amplification at high SNR.
% performs slightly worse at high SNR due to stronger noise amplification.

%Moreover, it can be seen that PB-S-SP loses its performance advantage over PB-EGC at high SNR. This can be attributed to the fact that noise amplification plays only a minor role. % at high SNR. %We note that the EGC and no-BF schemes employ a larger sum transmit power (i.e., $\sR \sP$ instead of $\sP$) 
 %P-SP is based on different relay sum powers. EGC employs a sum power of outperforms P-SP can be attributed to the fact that EGC may employs a larger sum relay transmit power (i.e., $\sR \sP$ instead of $\sP$). %The same statements hold when comparing EGC and PSP. 
%The reason that EGC outperforms P-SP can be attributed to the different relay sum powers. %the fact that EGC may employs a larger sum relay transmit power (i.e., $\sR \sP$ instead of $\sP$). %The same statements hold when comparing EGC and PSP. 

% the curves reveal that the SNR-based scheme PB-SSP allows to approach optimal power allocation of the relays. Therefore, PB-SSP shows a significant performance advantage in comparison to PB-PSP, which uses a power-based measure. A diversity of 2 can be measured for PB-SSP, SSP, and SIP which agrees with the observations in \cite{Jing:2007aa}. The reason that SIP outperforms SSP can be attributed to the fact that SIP may employ a larger aggregate relay transmit power than SSP. The same statements hold when comparing EGC and PSP. 

\subsection{Realistic Scenario}
We next use  
% a 2-relay system with $800$\,kHz bandwidth employing 
independent, time-varying flat fading channels with Jakes Doppler profile and the same path loss model as in Section \ref{ssec:ideal}. % (based on the model used in \cite{zemen_sp05}).  
Furthermore, the destination uses \eqref{eq:hest} and %, \eqref{eq:Prxest}, and 
\eqref{eq:snrest} to estimate 
% In this practical scenario we evaluate the performance measures and
the compound channel, the received signal power, and the instantaneous SNR.
%  according to \eqref{eq:hest}, \eqref{eq:Prxest}, and \eqref{eq:snrest} using a 
To this end, each transmission frame contains $|\sTpset|\!=\!10$ pilot symbols in addition to $|\sTdset|\!=\!40$ data 
symbols. 
% within a frame length of $62.5\,\upmu$s. 
The normalization in \eqref{eq:asi} is achieved by measuring the receive power at the relays during one frame.

\begin{figure}[t]
\centering
%\subfigure[]{\includegraphics[scale=0.37]{plots/plot_track_conv_R2_power.eps}}
%\hfill
%\subfigure[]{
\includegraphics[scale=0.37]{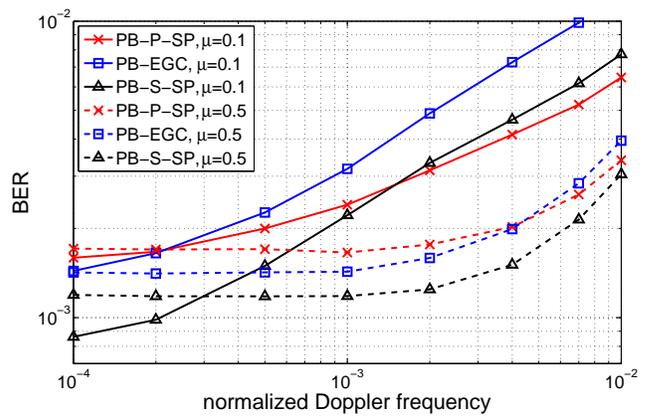}%}
\vspace*{-1mm}
%\hfill
%\subfigure[]{\includegraphics[scale=0.31]{plots/plot_ber_R2_sub.eps}}
%\subfigure[]{\includegraphics[scale=0.38]{plots/plot_cap_4x2_16QAM_all.eps}}
%\vspace*{-2mm}
\caption{BER versus normalized Doppler frequency for various PB-BF schemes using P/M perturbation under a realistic scenario ($\sPsum/\sNo\!=\!22$\,dB).}
%\vspace*{-1mm}
%\caption{Tracking scenario for $\sR\!=\!2$ relays: (a) tracking performance for PB-SSP (20km/h), and (b) BER versus velocity for various PB-BF schemes.}
\label{fig:tracking}
\vspace*{-2mm}
\end{figure}

We analyze the tracking capabilities of %beamforming using 
P/M perturbation (with $\sbeta\!=\!0.1$ and $\sbeta\!=\!0.5$) 
% by plotting 
in terms of BER versus normalized Doppler frequency (i.e., Doppler in Hertz times frame length in seconds) for 
$\sPsum/\sNo\!=\!22$\,dB (see Fig.\ \ref{fig:tracking}). %with adaption parameter $\sbeta=0.1$ and $\sbeta=0.3$
%In this scenario $\sDset$ moves with a velocity of $3$\,km/h (corresponds to Doppler frequency of $90$\,Hz) from... {\em not finished yet!!!}
In general, %it can be seen that the 
the BER degrades with increasing Doppler.
% (step size $\sbeta$ and frame length are fixed). 
At high Doppler frequencies the relay weights cannot be adapted fast enough to the channel variations (note that in practice, the feedback delay will add on top of this). Moreover, if there are channel variations within a frame, the compound channel and the objective function cannot be estimated accurately. 
Even at low Doppler, there is an order of magnitude BER penalty for PB-S-SP (cf.\ 
Fig.\ \ref{fig:tracking} with $\sbeta\!=\!0.1$ and Fig.\ \ref{fig:ideal}(c) at $\sPsum/\sNo\!=\!22$\,dB).

% However, performance can be tuned by adapting the step-size parameter. 
We observe that at low Doppler frequencies a small step size ($\sbeta\!=\!0.1$) performs better whereas at higher Doppler frequencies 
% BER performance tends to be more robust for 
a larger step size  ($\sbeta\!=\!0.5$) is advantageous since it allows quicker adjustment of the relay weights. 
% In fact, it can be seen in Fig.\ \ref{fig:tracking} that a small $\sbeta$ can result in a significant performance degradation at high Doppler frequencies; 
Note that with $\sbeta\!=\!0.1$, PB-S-SP looses its entire performance advantage over PB-P-SP at high Doppler frequencies. %and $\sbeta\!=\!0.3$ shows that there is a tradeoff between fast adjustment of the weights for high Doppler frequencies and  outperform the curves with $\sbeta\!=\!0.3$ at small Doppler frequencies, it can be seen that 
%In fact, a larger adaption parameter behaves more robust for high Doppler frequencies. 

In time-varying scenarios, T/R suffers from the fact that the beamforming weights are not 
updated when the channel quality gets worse. This can be circumvented by building a forgetting factor into the performance measure $\sfbmtwo$.

% ************** CONCLUSION *********************************
\section{Conclusion}
\label{sec:conclusion}
We have investigated scalable perturbation-based distributed beamforming protocols in wireless relay networks that exploit 1-bit feedback to %``learn'' 
approach the optimal beamforming weights in an adaptive manner
while avoiding CSI at the relay nodes. 
%We presented two different perturbation approaches, both of which 
We used a deterministic perturbation set to optimize either received signal power or SNR at the destination under per-relay or sum power constraints. %, and provided a scalable protocol. 
% It is shown that a distributive and scalable protocol can be realized using a deterministic perturbation set. 
% Furthermore, we showed that optimal relay power control can be approached when 
The best performance %(i.e., full diversity) 
was observed with SNR as objective function under a sum power constraint. At high SNR, equal gain combining appears to be preferable over power optimization under a sum-power constraint.
% non-uniform power allocation obtained by power optimization.
In time-varying environments, the proposed perturbation schemes require a careful choice of the step-size parameter and the transmission frame length.

%% use section* for acknowledgement
%\section*{Acknowledgment}


%\begin{thebibliography}{1}


%\end{thebibliography}

% BiBTeX files (here: strings, refs). The IEEEbib.bst bibliography
\bibliographystyle{IEEEtran}
\bibliography{GC08bib_arxiv}

\end{document}